\documentclass[final,9p,times]{elsarticle}
\usepackage[a4paper,margin=1in]{geometry}
\usepackage{setspace} 
\usepackage{graphicx}
\usepackage{amssymb}
\usepackage{newtxtext}
\usepackage{newtxmath}
\usepackage{natbib}
\usepackage{xcolor}
\usepackage{hyperref}
\usepackage{cleveref} 
\usepackage{float}
\usepackage{multirow}
\usepackage{amsmath}
\usepackage{booktabs}
\usepackage{tabularx}
\usepackage{array} 
\usepackage{caption}  
\usepackage{algorithm} 
\usepackage{algpseudocode}
\hypersetup{
    colorlinks=true,
    citecolor=blue,    
    linkcolor=blue,    
    filecolor=blue,    
    urlcolor=blue      
}

\crefname{equation}{equation}{Eqs.}
\Crefname{equation}{equation}{Eqs.}
\crefname{figure}{figure}{figures}
\Crefname{Figure}{Figure}{Figures}
\crefname{algorithm}{Algorithm}{Algorithms}
\Crefname{algorithm}{Algorithm}{Algorithms}
\crefname{section}{section}{sections}
\Crefname{Section}{Section}{Sections}
\crefname{table}{table}{tables}
\Crefname{Table}{Table}{Tables} 

\definecolor{hcorange}{RGB}{245,130,48}
\definecolor{hcnavy}{RGB}{0,0,128}
\definecolor{hcblue}{RGB}{0,130,200}
\definecolor{hcpink}{RGB}{250,190,190}
\definecolor{hcbrown}{RGB}{128,0,0}
\definecolor{hclavender}{RGB}{230,190,255}
\definecolor{hcgrey}{RGB}{128,128,128}
\definecolor{hcgreen}{RGB}{60,180,75}
\definecolor{hcred}{RGB}{230, 25, 75}
\definecolor{darkred}{RGB}{128,0,0}
\definecolor{darkblue}{RGB}{0,0,170}
\definecolor{hcblack}{RGB}{0,0,0}
\definecolor{hcred2}{RGB}{255,0,0}

\PassOptionsToPackage{normalem}{ulem}
\usepackage{ulem}

\newcommand{\rewrite}[1]{{\color{black}{}#1}} 

\begin{document}

\begin{frontmatter}

\title{State-augmented deep reinforcement learning for active flow control around an elliptical cylinder
} 

\author[]{Wang Jia}

\author[]{Hang Xu}
\ead{hangxu@sjtu.edu.cn}
\affiliation{organization={State Key Laboratory of Ocean Engineering, School of Ocean and Civil Engineering, Shanghai Jiao Tong University},
            city={Shanghai},
            postcode={200240}, 
            country={China}}
\begin{abstract}

We employ reinforcement learning algorithms to develop control strategies for manipulating synthetic jets on an elliptical cylinder, targeting robust flow control across varying aspect ratios ($Ar$). Our results show that for high aspect ratios ($Ar = 1$ and $0.75$), the learned strategies effectively mitigate vortex shedding, achieving drag reductions of 13.4\% and 20.1\%, respectively. As the $Ar$ decreases, more complex wake dynamics challenge flow control. For $Ar = 0.5$ and $0.25$, drag reductions of 26.88\% and 30.37\% are achieved, though vortex suppression becomes increasingly difficult. For extremely slender cylinders ($Ar = 0.1$), we enhance control performance by enriching dynamic feature representation and using Kalman filtering for state estimation. Unlike conventional methods, our approach successfully achieves complete vortex shedding suppression, a maximum drag reduction of 32.2\%, and eliminates periodic oscillations, stabilizing the flow. These results demonstrate the efficacy of state-enhanced RL in controlling highly slender geometries.

\end{abstract}


\end{frontmatter}

\section{INTRODUCTION}

\begin{figure}
    \centering
    \includegraphics[width=0.95\textwidth]{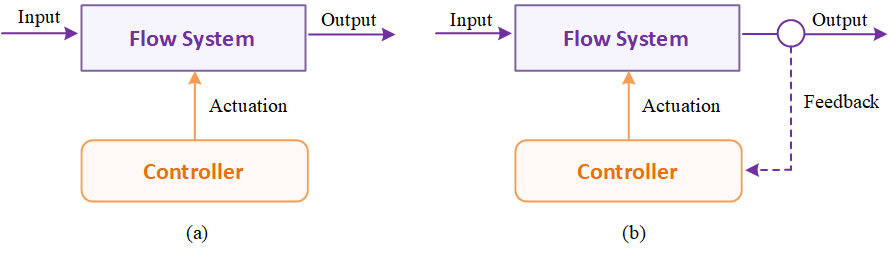}
    \caption{Comparison of open-loop and closed-loop control strategies. (a) Open-loop control applies a predetermined input without feedback, while (b) closed-loop control adjusts the control input based on real-time feedback from the flow system.}
\label{fig:figure1}
\end{figure}

Flow control technology aims to optimize flow structures through active or passive strategies to achieve key objectives such as drag reduction, vibration suppression, and enhanced heat transfer, thereby improving overall system performance and energy efficiency \cite{SCOTTCOLLIS2004237}. 
These strategies are typically divided into two categories: open-loop and closed-loop approaches \cite{jahanmiri2010active}. Open-loop control operates without feedback, where control actions are predetermined and remain fixed, regardless of the system’s state \cite{Collins1993}. In contrast, closed-loop control incorporates real-time feedback from the system, allowing for continuous adjustments to control actions, which facilitates dynamic adaptation to changes in the flow conditions \cite{brunton2015closed}. The distinction between these two approaches is illustrated in figure \ref{fig:figure1}.
Closed-loop control is particularly valuable due to its ability to utilize real-time feedback mechanisms, which significantly enhance flow performance and enable adaptation to dynamic conditions \cite{annurevCattafesta, BEWLEY200121}. This makes it especially effective in managing complex flow structures. However, despite its advantages, implementing closed-loop control presents significant technical and practical challenges \cite{annurevchoi}. These challenges stem from the need for advanced hardware, such as high-precision sensors and fast-response actuators, as well as sophisticated software to ensure effective operation \cite{Bossanyi, HE2025126300}. 

While traditional control methods have been widely used, they often struggle with complex and dynamic flow conditions, particularly in cases involving highly slender geometries and extreme flow conditions. 
In recent years, the rapid advancements in machine learning have revolutionized various scientific and engineering disciplines, including fluid mechanics \cite{Viquerat,GARNIER,rabault2020deep}. Through the use of its powerful data-driven capabilities, machine learning has emerged as a transformative tool for tackling complex flow problems that are difficult to address using traditional methods \cite{brunton2015closed,ren2020active}. In particular, ML techniques have been applied across a wide range of fluid mechanics areas, including turbulence modeling, flow control, optimization, and data assimilation \cite{annurevfluid,LI2022100849,subgridscale,dong2024interactive,KIM2022111263,BHOLA2023112018}.
For instance, supervised learning algorithms have been employed to develop accurate surrogate models for computationally expensive simulations, while unsupervised learning methods have been used to identify and analyze flow patterns. Reinforcement learning (RL), in particular, has shown immense potential in active flow control by enabling adaptive strategies for drag reduction, lift suppression, and vortex shedding mitigation \cite{Viquerat,GARNIER,ren2020active}.
These applications highlight how machine learning can extract valuable insights from high-dimensional flow data and offer efficient solutions to nonlinear and multiscale problems \cite{brunton2015closed,Viquerat}.

Among various machine learning techniques, deep reinforcement learning (DRL) has emerged as a promising approach for active flow control due to its ability to learn optimal control strategies through trial and error in complex environments. 
Since the pioneering work by \citeauthor{rabault2019artificial} \cite{rabault2019artificial}, the rapid development of DRL-based AFC techniques has been further facilitated by the open-sourcing of codes from several studies \cite{rabault2019artificial,liReinforcementlearning,Xia2024,Dixiapnas,Dynamic2024,wangDRLinFluids}, enabling researchers in CFD and DRL to quickly enter this emerging field. In particular, \citeauthor{salehi2024efficient} \cite{salehi2024efficient} presented a TensorFlow-based intrusive DRL-CFD framework that eliminates external data exchange while leveraging message-passing interfaces to parallelize computationally intensive CFD cases. Meanwhile, the study by \citeauthor{weiner2024model} \cite{weiner2024model} demonstrated that a model-based learning approach can reduce overall training time by up to 85\% for the fluidic pinball test case. Additionally, \citeauthor{dong2024interactive} \cite{dong2024interactive} proposed a novel DRL-based wind tunnel testing platform that manages interactions across mainstream communication protocols and integrates reinforcement learning libraries, significantly reducing the interaction cost between DRL algorithms and wind tunnel tests. 
Furthermore, \citeauthor{jia2024jetsactuator} \cite{jia2024jetsactuator} found that placing synthetic jets at primary flow separation points achieves optimal control, while the data-driven active control strategy developed by \citeauthor{jia2024robust} \cite{jia2024robust} achieved drag reductions of approximately  14.4\%, 26.4\%, 38.9\%, and 47.0\%  for $Re = 100$, 200, 300, and 400, respectively, successfully suppressing vortex shedding and achieving optimal control performance.

Despite the significant progress in DRL-based flow control, challenges remain, especially for highly slender geometries such as elliptical cylinders. 
These geometries present complex wake dynamics and multi-stage vortex shedding transitions, making it difficult to achieve complete vortex shedding suppression and stable flow control. Although the powerful decision-making capability of DRL has been demonstrated across a wide range of flow control applications, effective control of flows around highly slender elliptical cylinders remains a particularly challenging research problem.
In this context, we aim to explore whether DRL-based AFC can completely suppress vortex shedding for such cylinders, which is essential for extending DRL applications to complex geometries and extreme flow conditions.

Anothe critical challenge in this area is balancing flow stability with energy efficiency during the control process. To address this, we enhance the foundational DRL training framework to manage the complexities associated with extreme aspect ratios ($Ar$). We then evaluate the effectiveness of the DRL-based AFC on two-dimensional (2D)elliptical cylinders with $Ar$ of 0.25, 0.5, 0.75, and 1. The objective is to achieve complete vortex shedding suppression while simultaneously meeting the dual demands of stability and energy efficiency under highly challenging flow conditions, all while maintaining optimal control performance.
Our study targets a key gap in the literature on DRL-based flow control for highly slender geometries, in which traditional methods often prove inadequate for suppressing unsteady wake dynamics.
The main contributions of this work are summarized as follows:
\begin{itemize}
    \item A comprehensive evaluation of vortex shedding suppression across a range of elliptical cylinder aspect ratios, thereby revealing the performance boundaries of DRL in handling complex, transitional wake flows.
    \item An enhanced observation-state formulation that improves the learning efficiency and control effectiveness of DRL algorithms, enabling more robust active flow control.
    \item A quantitative analysis of the trade-off between control energy consumption and flow stabilization, providing valuable insights into the development of energy-efficient DRL-based control strategies.
\end{itemize}

Overall, this work contributes to the broader objective of extending DRL applications to more geometrically complex and dynamically unstable flow configurations, thus advancing the practical potential of intelligent flow control in real-world engineering scenarios.
This paper is structured as follows. Section~\ref{sec:Methodology} describes the numerical simulation setup and the DRL framework employed for active flow control. Section~\ref{sec:Results} presents the training outcomes and evaluates the performance of the proposed control strategies. Finally, Section~\ref{sec:Conclusions} summarizes the key findings and outlines directions for future research.

\section{METHODOLOGY}\label{sec:Methodology}

\subsection{Flow Configuration}\label{sec:Simulation}

The geometry of elliptical cylinders with varying aspect ratios is considered, with each cylinder characterized by a major axis of $2b$ and a minor axis of $2a$. The center of each ellipse is denoted as point $O$, and the characteristic height of the cylinder is defined as $D = 2a$. The aspect ratio $Ar$, given by $Ar = a/b$, is used to examine four distinct configurations: a circular shape ($Ar = 1$) and progressively more slender forms ($Ar = 0.75, 0.5, 0.25, 0.1$). This range of aspect ratios enables a comprehensive investigation of the effect of cylinder shape on fluid dynamics.

\begin{figure}
    \centering
    \includegraphics[width=0.8\textwidth]{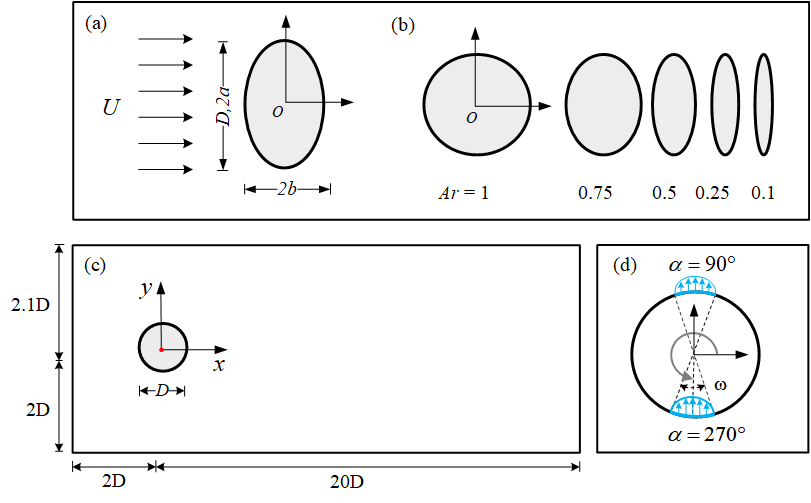}
    \caption{Schematic diagram of the physical model of flow around a two-dimensional elliptical cylinder. (a) Schematic of an incoming uniform flow with velocity $U$ impinging on an elliptical cylinder of major axis $2b$ and minor axis $2a$, where the characteristic length is defined as $D = 2a$. (b) Illustration of elliptical cylinders with different aspect ratios $Ar = a/b$, ranging from 1 (circular cylinder) to 0.1 (slender ellipse). (c) Computational domain configuration for flow simulations. The domain size is $20D \times 4.1D$. (d) Synthetic jets positioned at angular locations $\alpha = 90^\circ$ and $\alpha = 270^\circ$.}
\label{fig:figure2}
\end{figure}

The Navier-Stokes equations for an incompressible, viscous fluid within a domain $\Omega \subset \mathbb{R}^{nd}$ over a time interval $(0, T)$ form the fundamental framework for the current fluid dynamics analysis, expressed as follows:
\begin{subequations}
\begin{equation}
    \frac{\partial \mathbf{u}}{\partial t} + \mathbf{u} \cdot \nabla \mathbf{u} = -\nabla p + \frac{1}{Re} \nabla^2 \mathbf{u},  \quad \text{in} \quad \Omega \times (0, T),
\end{equation}
\begin{equation}
    \nabla \cdot \mathbf{u} = 0, \quad \text{in} \quad \Omega \times (0, T),
\end{equation}
\end{subequations}
where $\mathbf{u} = \mathbf{u}(\mathbf{x}, t)$ denotes the velocity field, $\mathbf{x}$ represents spatial coordinates, and $t$ is time. The pressure field is represented by $p = p(\mathbf{x}, t)$, and the Reynolds number $Re$ characterizes the ratio of inertial to viscous forces in the flow, defined as:
\begin{equation}
Re = \frac{\overline{U}L}{\nu},
\end{equation}
with $\overline{U}$ as the characteristic velocity at the inlet, $L$ as the characteristic length scale (equal to the cylinder diameter $D$ for flow around a cylinder), and $\nu$ representing the kinematic viscosity. We focus on comparing the drag coefficient $C_D$ and lift coefficient $C_L$, defined as:
\begin{align}
     C_D = \frac{2F_D}{\rho \overline{U}^2D}, \quad C_L = \frac{2F_L}{\rho \overline{U}^2D},
\end{align}
where $F_L$ and $F_D$ are the lift and drag forces integrated over the cylinder's surface, and $\rho$ is the fluid density. 
The numerical simulations are conducted using the open-source \texttt{OpenFOAM} platform, with the computational domain discretized via the finite volume method. The \texttt{pimpleFoam} solver, which integrates the PISO and SIMPLE algorithms, is employed to efficiently handle transient incompressible flows. Details of the Numerical scheme and the grid independence study are provided in \hyperref[app1]{Appendix I}.

The synthetic jet boundary conditions are designed to ensure smooth and controlled modulation of the jet velocity along the surface of the cylinder. Parameterized by the azimuthal angle $\alpha$ and the mass flow rate $Q_i$, these conditions dictate the jet's intensity and location.
\begin{equation}
f_{Q_i} = u_{\text{jet}}(\alpha; Q_i),
\end{equation}
For a jet on a cylinder with diameter $D$, centered at an azimuthal angle $\alpha_0$ and having a width $\omega$, the radial velocity profile is given by 
\begin{equation}
u_{\text{jet}}(\alpha, Q_i) = \frac{\pi}{\omega D} Q_i \cos\left(\frac{\pi}{\omega} (\alpha - \alpha_0)\right),
\end{equation}
with $\omega = 10^\circ$ defining the angular extent of the jet's influence. This setup ensures a symmetric velocity distribution around $\alpha_0$. For each geometry, there are two jets located at $\alpha_0 = 90^{\circ}$ and $270^{\circ}$, respectively. The mass flow rate $Q_i$ (where $i = 1, 2$ for two jets) governs the jet intensity, with higher $Q_i$ values corresponding to stronger velocities. Each synthetic jet alternates between blowing and suction phases, creating a periodic modulation essential for effective control performance. To satisfy mass conservation, the net mass flow rate from the two jets is kept at zero, ensuring that $Q_1 + Q_2 = 0$ and preventing the synthetic jet system from introducing any net mass flux into the domain, thus the total mass in the control volume. 

\subsection{Deep Reinforcement Learning Algorithm}\label{sec:drl}

\begin{figure}
    \centering
    \includegraphics[width=0.95\textwidth]{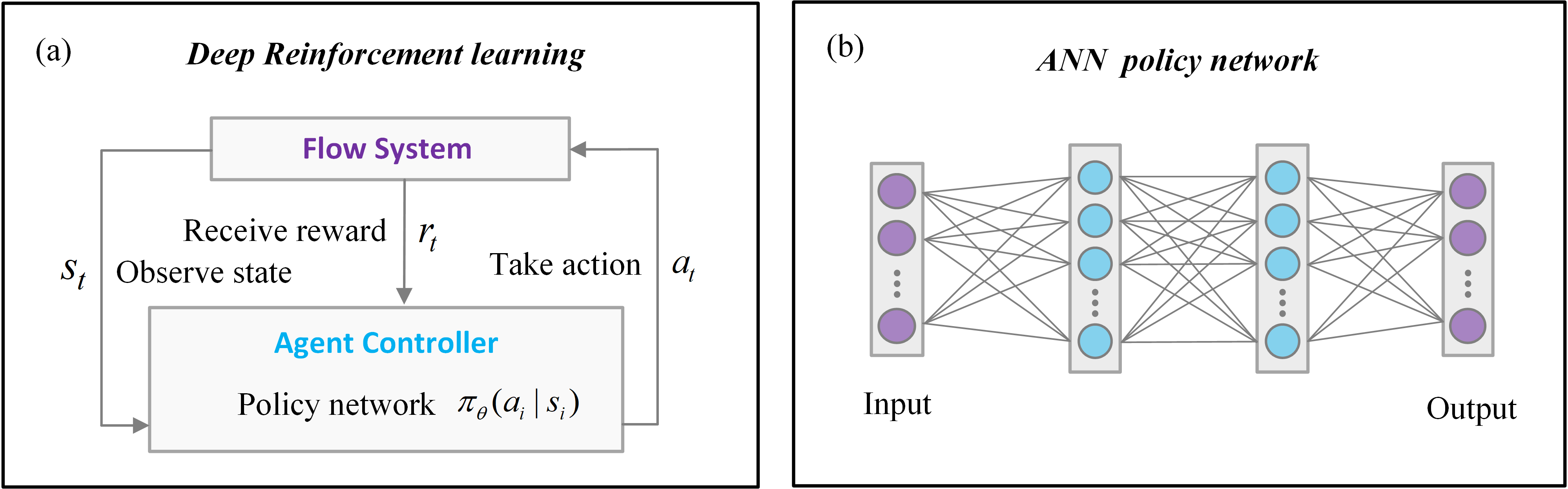}
    \caption{(a) Deep reinforcement learning framework; (b) ANN policy network architecture.}
\label{fig:figure3}
\end{figure}

Reinforcement Learning (RL) is a branch of machine learning focused on sequential decision-making, formalized as a Markov Decision Process (MDP) represented by the tuple $(S, A, P, R, \gamma)$. Here, $S$ denotes the state space, $A$ the action space, $P$ the state transition probability dictating the likelihood of moving to the next state $s_{t+1}$ given the current state $s_t$ and action $a_t$, $R$ is the reward function that assigns a scalar reward $r_t$ for the pair $(s_t, a_t)$, and $\gamma \in [0, 1)$ is the discount factor balancing immediate and future rewards. 
The objective of RL is to learn an optimal policy $\pi^*$ that maximizes the expected discounted cumulative reward through interactions between the agent and the environment, as illustrated in figure \ref{fig:figure3}.

Policy gradient methods are a class of RL algorithms that directly optimize the policy of an agent, rather than learning a value function (as in Q-learning methods).
Policy gradient methods optimize the policy $\pi_\theta(a|s)$ with parameters $\theta$ to maximize the expected cumulative reward:
\begin{equation}
J(\theta) = \mathbb{E}_\pi \left[ \sum_{t=0}^\infty \gamma^t R_{t+1} \right].
\end{equation}
The gradient of this objective is expressed as:
\begin{equation}
\nabla_\theta J(\theta) = \mathbb{E}_\pi \left[ \nabla_\theta \log \pi_\theta(a|s) Q^\pi(s, a) \right],
\end{equation}
where $Q^\pi(s, a)$ is the action-value function used to iteratively update $\theta$. However, directly applying policy gradient methods can lead to instability due to large policy parameter updates. Proximal Policy Optimization (PPO) addresses this by introducing constraints on policy updates to limit deviations between new and old policies, ensuring stable learning and improved convergence. The pseudocode for PPO with a clipped objective is as follows:
\begin{algorithm}
\caption{PPO with Clipped Objective}
\begin{algorithmic}[1]
\Require Initial policy parameters $\theta_0$, clipping threshold $\epsilon$
\For{$k = 0, 1, 2, \dots$}
    \State Collect set of partial trajectories $\mathcal{D}_k$ on policy $\pi_k = \pi(\theta_k)$
    \State Estimate advantages $\hat{A}_t^{\pi_k}$ using any advantage estimation algorithm
    \State Compute policy update
    \begin{equation}
    \theta_{k+1} = \arg \max_\theta \mathcal{L}^{\text{CLIP}}_{\theta_k}(\theta)
    \end{equation}
    by taking $K$ steps of minibatch SGD (via Adam), where
    \begin{equation}
    \mathcal{L}^{\text{CLIP}}_{\theta_k}(\theta) = \mathbb{E}_{\tau \sim \pi_k}\left[\sum_{t=0}^T \min \left(r_t(\theta)\hat{A}_t^{\pi_k}, \text{clip}\left(r_t(\theta), 1 - \epsilon, 1 + \epsilon\right)\hat{A}_t^{\pi_k} \right) \right]
    \end{equation}
\EndFor
\end{algorithmic}
\end{algorithm}

\begin{figure}
    \centering
    \includegraphics[width=\textwidth]{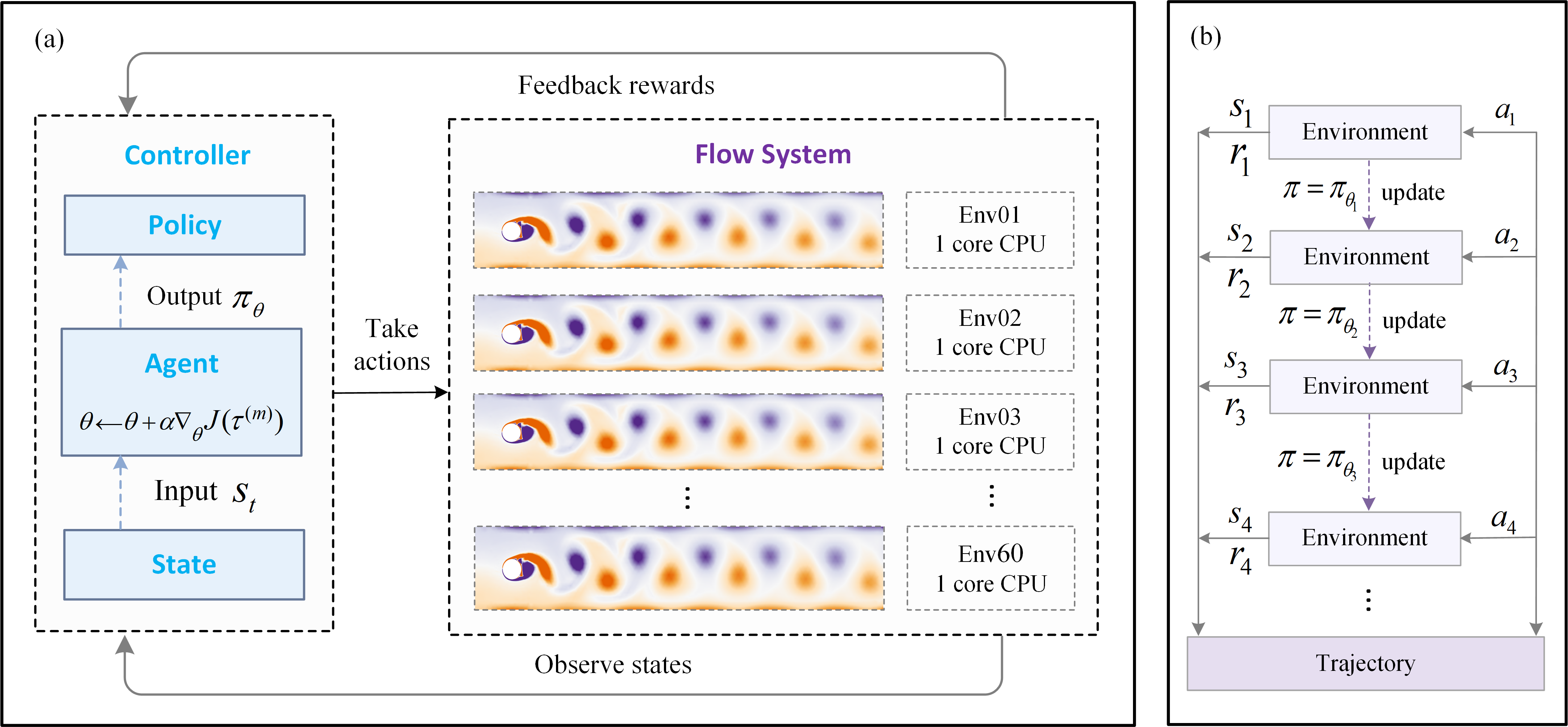}
    \caption{schematic diagram of the deep reinforcement learning framework for flow control in fluid system.}
\label{fig:figure4}
\end{figure}

\subsection{Framework of Active Flow Control with DRL}\label{subsec:framework}

We implement DRL for AFC using a PPO agent that interacts with a computational fluid dynamics (CFD)-based environment, as illustrated in figure \ref{fig:figure4}. The agent gathers critical flow field information through strategically positioned probes, which capture key flow characteristics around the cylinder and its wake. This real-time feedback enables the agent to dynamically adjust the velocity of synthetic jets by modulating their unit mass flow rate.  
At each time step $t$, the agent assesses the current state $s_t$, selects an action $a_t$ based on the policy $\pi$, and executes the corresponding control action on the synthetic jets.
This action influences the flow field, resulting in a corresponding reward $r_t$, which quantifies the effectiveness of the control decision. Through continuous interaction with the CFD environment, the agent refines its control strategy iteratively.
The reward function is designed to promote drag reduction while suppressing lift fluctuations. In this work, the reward function is defined as 
\begin{eqnarray}
r_{T_i} = C_{D,0} - \left(C_D\right)_{T_i} - 0.1 \left| \left(C_L\right)_{T_i} \right|,
\end{eqnarray}  
where $ C_D $ and $ C_L $ denote the drag and lift coefficients, respectively, while $ C_{D,0} $ represents the baseline drag coefficient. A penality parameter with value 0.1 is added to mitigate the fluctuation in lift.

\subsection{Enhanced Active Flow Control framework}\label{Enhanced}

The effectiveness of DRL in controlling complex dynamical systems depends on the quality of state information provided to the agent. Conventional state observations, limited in dimensionality and temporal resolution, often hinder accurate interpretation of flow dynamics. To address this, we enhance state representations by integrating one-dimensional, two-dimensional, and filtered data, offering a more comprehensive depiction of the underlying physics. This approach captures intricate spatiotemporal dynamics, improving the agent’s perception. To further optimize control performance, we utilize three distinct types of state information.

\begin{enumerate}
    \item \textbf{One-Dimensional State Information:} The state space, denoted as $\mathbf{z}_{\text{1D}} \in \mathbb{R}^n$, is a one-dimensional vector that contains physical information gathered from $n$ observation probes at the final time step of each control cycle $t_c$,
    \[
    \mathbf{z}_{\text{1D}} = \left[ z_{t_c,1}, z_{t_c,2}, \ldots, z_{t_c,n} \right]^T,
    \]
    where $z_{t_c,i}$ represents the observed physical quantity (e.g., velocity, pressure) at probe $i$. This compact representation serves as input to the artificial neural network within the DRL framework.

    \item \textbf{Two-Dimensional State Information:} In an attempt to enrich the state space, we encode temporal information by providing a two-dimensional tensor. This enriched state space, represented as $\mathbf{Z}_{\text{2D}} \in \mathbb{R}^{n \times k}$, comprises physical information collected from $n$ observation probes over $k$ time steps within each control cycle,
    \[
    \mathbf{Z}_{\text{2D}} = \begin{bmatrix} 
    z_{t_1,1} & z_{t_2,1} & \cdots & z_{t_k,1} \\
    z_{t_1,2} & z_{t_2,2} & \cdots & z_{t_k,2} \\
    \vdots & \vdots & \ddots & \vdots \\
    z_{t_1,n} & z_{t_2,n} & \cdots & z_{t_k,n} 
    \end{bmatrix},
    \]
    where $z_{t_j,i}$ indicates the observed quantity at observation point $i$ at time step $t_j$. This representation effectively captures the spatial and temporal dynamics of the flow.

    \item \textbf{Enhanced Two-Dimensional State Information with Kalman Filtering:} This type, denoted as $\mathbf{Z}_{\text{stacked}} \in \mathbb{R}^{k \times (n + m)}$.
    The Kalman Filter (KF) is an optimal recursive estimation algorithm for extracting true system states in high-noise environments \cite{6400245}. We employs a Real-Time Kalman Filter to refine time-series data, constructing an Enhanced State that integrates raw observations with filtered estimates, thereby improving state representation and robustness.
    The enhanced feature matrix is defined as:
    \[
    \mathbf{Z}_{\text{stacked}} = \begin{bmatrix} 
    \mathbf{z}_1 & \mathbf{x}_{1|1} \\ 
    \mathbf{z}_2 & \mathbf{x}_{2|2} \\ 
    \vdots & \vdots \\ 
    \mathbf{z}_T & \mathbf{x}_{T|T} 
    \end{bmatrix}
    = \begin{bmatrix} 
    \mathbf{z}_t & \mathbf{x}_{t|t} 
    \end{bmatrix}_{t=1}^T,
    \]
    where $\mathbf{z}_t \in \mathbb{R}^n$ represents the original observation at time step $t$, and $\mathbf{x}_{t|t} \in \mathbb{R}^m$ denotes the state estimate filtered by the Kalman filter. This enhanced two-dimensional state representation integrates spatial-temporal flow features while mitigating noise.
    Details related to the implementation of the Kalman filter are provided in \hyperref[app2]{Appendix II}.
\end{enumerate}

\section{RESULTS}\label{sec:Results}

In section~\ref{sec:sec1}, we present a comprehensive analysis of flow control effectiveness across elliptical cylinders with varying aspect ratios ($Ar = 1, 0.75, 0.5, 0.25$), identifying key trends in the DRL agent’s performance as cylinder slenderness increases in order to assess the robustness and limitations of the proposed model.
Building upon these findings, section~\ref{sec:sec2} investigates the suppression of vortex shedding in an extremely slender elliptical cylinder ($ Ar = 0.1 $), where we refine the DRL-based control framework to enhance its effectiveness in mitigating vortex shedding.

\subsection{Effectiveness of flow control for elliptical cylinders with $Ar = 1, 0.75, 0.5, 0.25$}\label{sec:sec1}

To assess the effectiveness of DRL in active flow control, we investigate the performance of DRL-based control strategies for elliptical cylinders with varying $ Ar $ at a Reynolds number of 100. As described in \ref{subsec:framework}, our evaluation focuse on several key metrics, including drag reduction, suppression of lift fluctuations, elimination of vortex shedding, and actuation energy consumption.

\begin{figure}[hbt!]
    \centering
    \includegraphics[width=\textwidth]{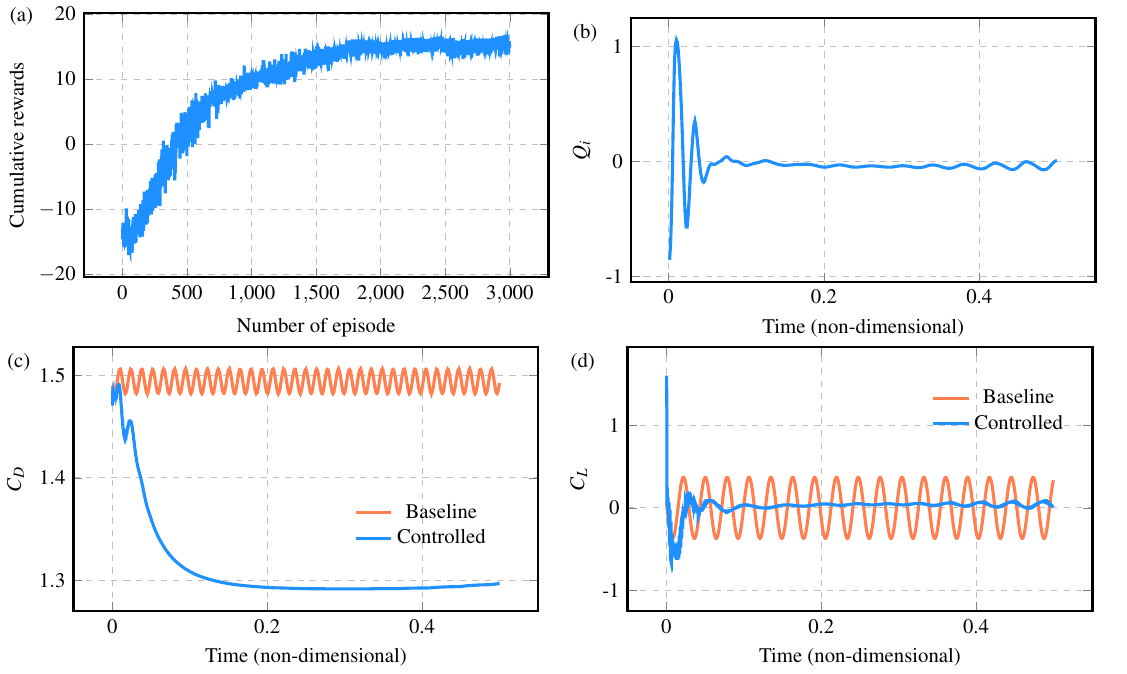}
    \caption{Performance analysis of flow control strategy for flow around a cylinder with $Ar=1$ at $Re=100$. (a) Cumulative rewards over training episodes. (b) Jet mass flow rate variation with time. (c) Time history of drag coefficient $C_D$ for baseline and controlled cases. (d) Time history of lift coefficient $C_L$ for baseline and controlled cases.}
\label{fig:figure5}
\end{figure}

In figure~\ref{fig:figure5}, the DRL training results for an elliptical cylinder with $ Ar = 1 $ are illustrated.
The cumulative reward function increases significantly throughout the DRL training process and eventually stabilizes, indicating the successful achievement of the desired objectives, as shown in figure~\ref{fig:figure5}(a).
The mass flow rate of the synthetic jets exhibits pronounced oscillations during the initial phase of control but rapidly stabilizes near its minimum value, highlighting the energy efficiency of the proposed control strategy, as shown in figure~\ref{fig:figure5}(b).
Furthermore, the periodic oscillations of the drag coefficient $C_D$ are completely eliminated. 
At the initial stage, the $C_D$ of the elliptical cylinder undergoes a rapid decline before quickly stabilizing.
Ultimately, this abrupt reduction results in a 13.41\% decrease in $C_D$, as shown in figure~\ref{fig:figure5}(c).
Additionally, the periodic oscillations of the lift coefficient $C_L$ in the baseline flow 
are suppressed by 99\% under flow control. 
This further confirms the effectiveness of the proposed control strategy in stabilizing the flow and mitigating unsteady aerodynamic forces, as evidenced by the results presented in figure~\ref{fig:figure5}(d).

\begin{figure}[hbt!]
    \centering
    \includegraphics[width=\textwidth]{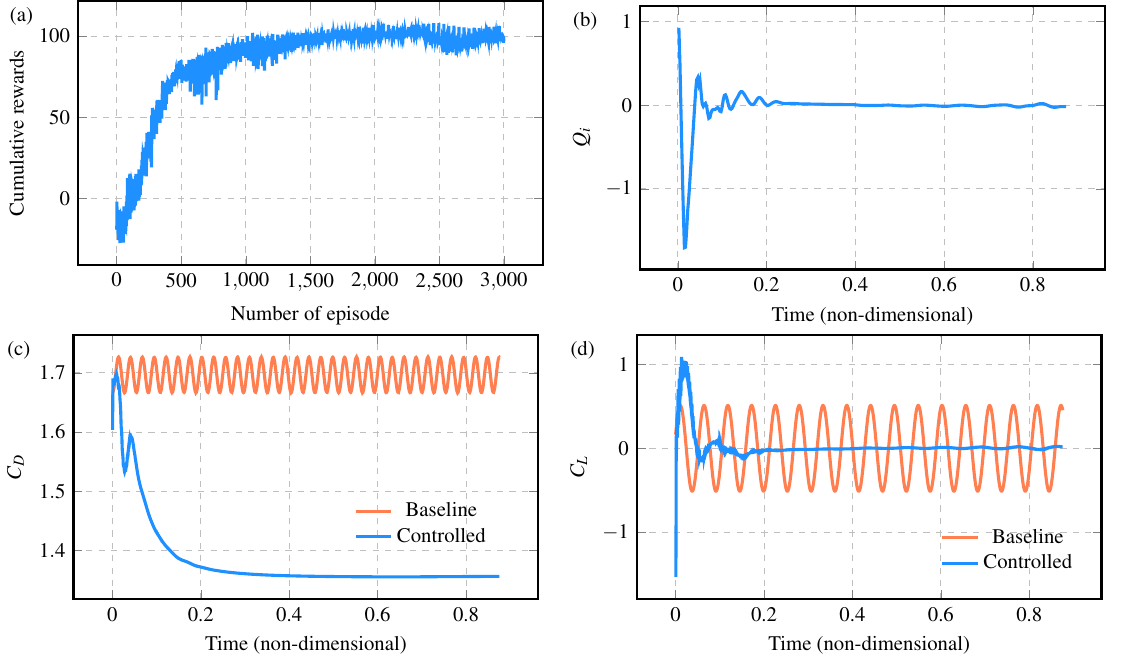}
    \caption{Performance analysis of flow control strategy for flow around a cylinder with $Ar=0.75$ at $Re=100$. (a) Cumulative rewards over training episodes. (b) Jet mass flow rate variation with time. (c) Time history of drag coefficient $C_D$ for baseline and controlled cases. (d) Time history of lift coefficient $C_L$ for baseline and controlled cases.}
\label{fig:figure6}
\end{figure}

\begin{figure}
    \centering
    \includegraphics[width=\textwidth]{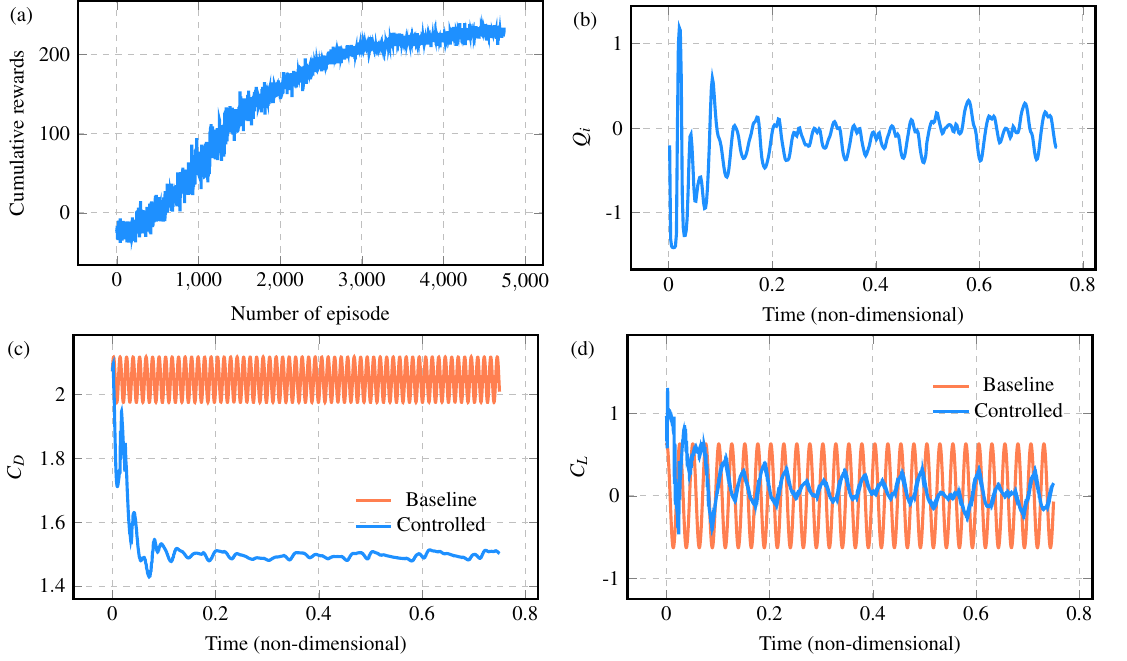}
    \caption{Performance analysis of flow control strategy for flow around a cylinder with $Ar=0.5$ at $Re=100$. (a) Cumulative rewards over training episodes. (b) Jet mass flow rate variation with time. (c) Time history of drag coefficient $C_D$ for baseline and controlled cases. (d) Time history of lift coefficient $C_L$ for baseline and controlled cases.}
\label{fig:figure7}
\end{figure}

\begin{figure}
    \centering
    \includegraphics[width=\textwidth]{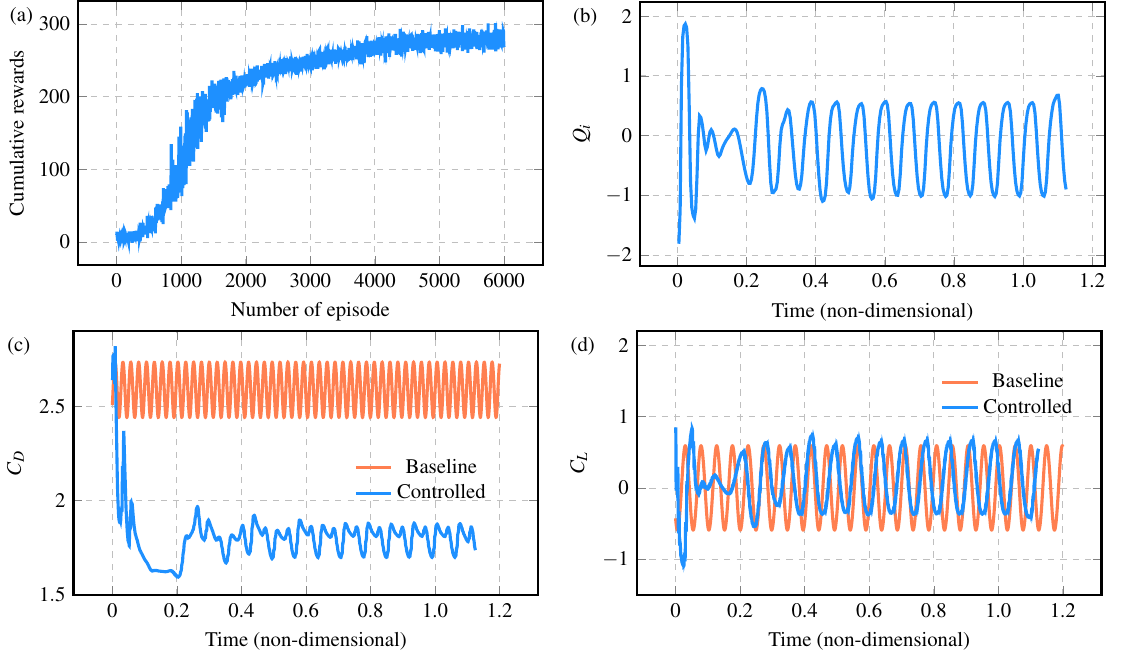}
    \caption{Performance analysis of flow control strategy for flow around a cylinder with $Ar=0.25$ at $Re=100$. (a) Cumulative rewards over training episodes. (b) Jet mass flow rate variation with time. (c) Time history of drag coefficient $C_D$ for baseline and controlled cases. (d) Time history of lift coefficient $C_L$ for baseline and controlled cases.}
\label{fig:figure8}
\end{figure}

The DRL training outcomes for the elliptical cylinder with $ Ar = 0.75 $ are illustrated in figure~\ref{fig:figure6}.
The various quantities of interest exhibit a strong resemblance to the trends observed for the elliptical cylinder with $ Ar = 1 $, achieving effective drag reduction and lift fluctuation suppression.  
Moreover, the temporal dynamics of the control strategy, reflected in the modulation of external energy input, follow a similar trajectory to that of the elliptical cylinder with $ Ar = 1 $.
Notably, the DRL-based control framework seamlessly adapts to the modified geometry without requiring any hyper-parameter tuning, underscoring its intrinsic robustness and generalizability in active flow control applications.

However, the same model encounters challenges in obtaining an effective control strategy for elliptical cylinders with $ Ar = 0.5 $ and $ Ar = 0.25 $, as shown in figure~\ref{fig:figure7} and figure~\ref{fig:figure8}.
A significantly greater number of episodes is required for the cumulative reward function to converge, as evidenced by the 4800 episodes needed for the elliptical cylinder with $ Ar = 0.5 $ and 6000 episodes for the elliptical cylinder with $ Ar = 0.25 $.
Although a significant reduction in the drag coefficient and suppression of the lift coefficient are observed in both cases, the controlled flow still exhibits periodic oscillations in both drag and lift coefficients. This indicates that the stabilization effect of the control strategy remains limited, which is further reflected in the fluctuations of the synthetic jet mass flow rate. These findings highlight the limitations of the current control strategy in achieving fully stabilized flow dynamics under these conditions.

\begin{figure}[hbt!]
    \centering
    \includegraphics[width=\textwidth]{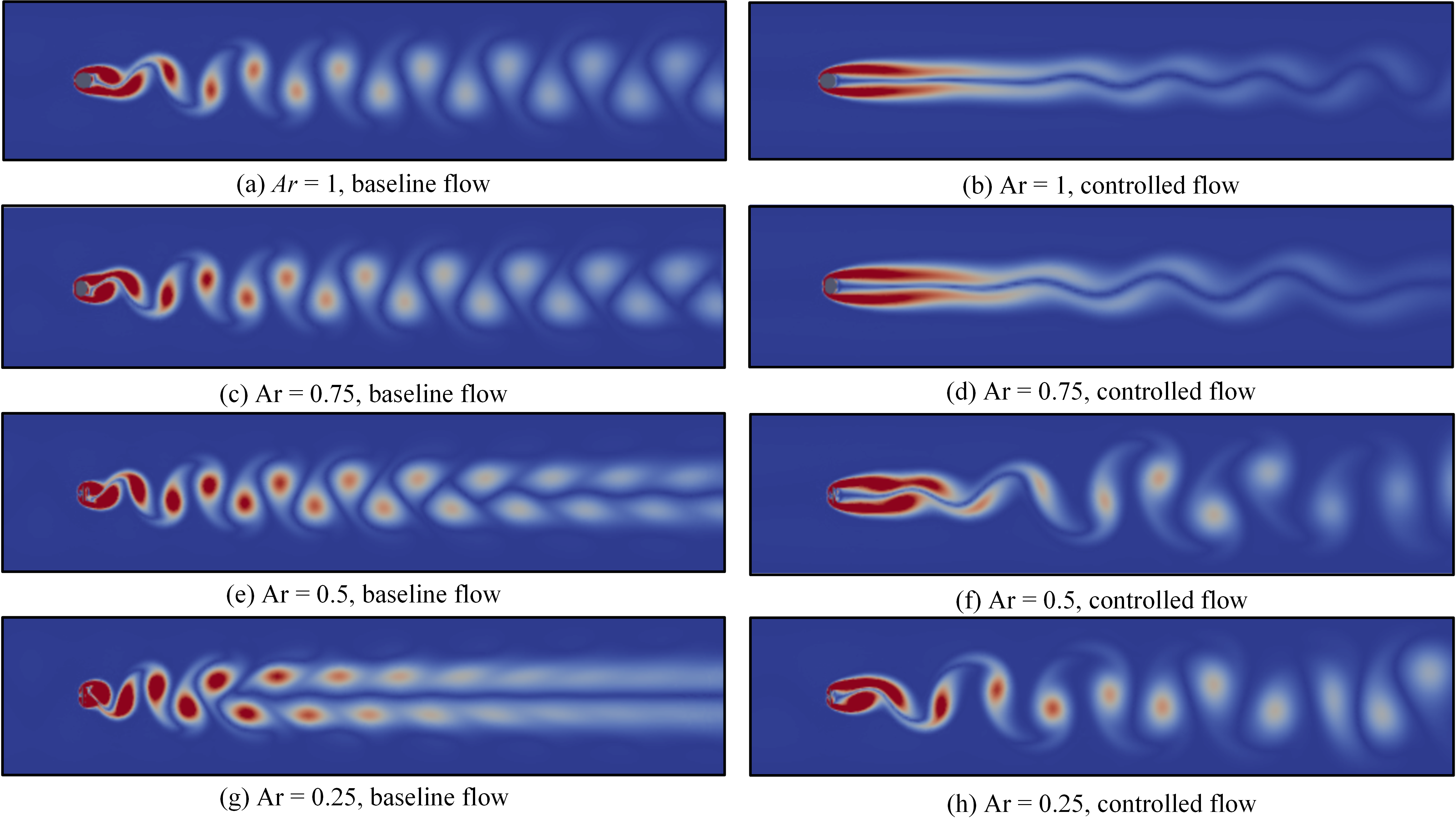}
    \caption{The instantaneous vorticity field of the elliptical cylinder before and after control. (a) $Ar =1$, baseline flow; (b) $Ar =1$, controlled flow; (c) $Ar =0.75$, baseline flow; (d) $Ar =0.75$, controlled flow; (e) $Ar =0.5$, baseline flow; (f) $Ar =0.5$, controlled flow; (g) $Ar =0.25$, baseline flow; (h) $Ar =0.25$, controlled flow.      
    }
\label{fig:figure9}
\end{figure}

\begin{figure}[hbt!]
    \centering
    \includegraphics[width=\textwidth]{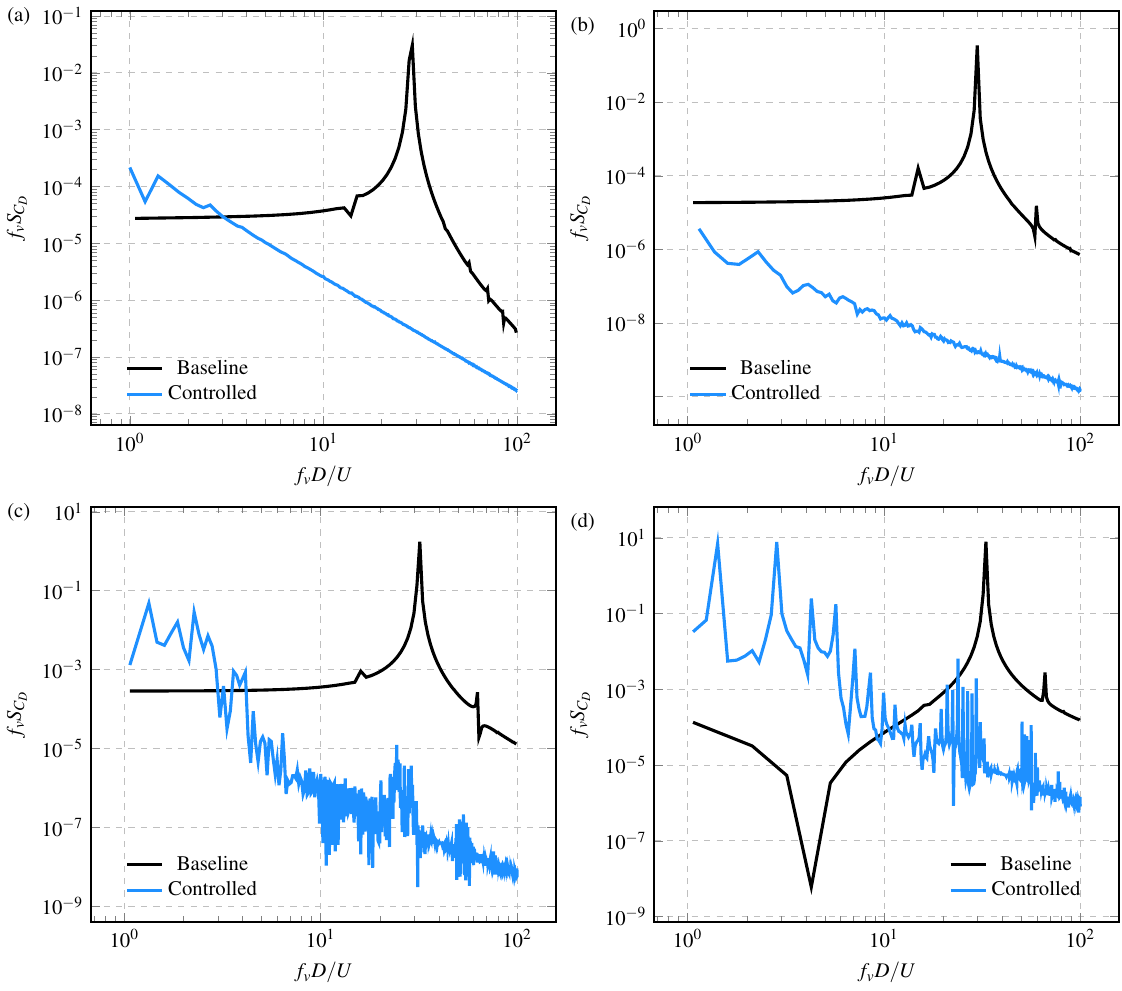}
    \caption{Power spectral density of the drag coefficient $C_D$ during the period of non-dimensional time, highlighting the differences in wake dynamics between baseline (uncontrolled) and controlled flows across aspect ratios. (a) $Ar = 1$; (b) $Ar = 0.75$; (c) $Ar = 0.5$; (d) $Ar = 0.25$.}
    \label{fig:figure10}
\end{figure}

To further elucidate the differences in control effectiveness, we conducted a qualitative comparison of the flow field shown in figure~\ref{fig:figure9} and a quantitative analysis of the power spectral density (PSD) of the drag coefficient presented in figure~\ref{fig:figure10}.  
The vorticity fields for various $ Ar $ values reveal distinct vortex shedding patterns in the wake. At $ Ar $ = 1 and $ Ar $ = 0.75, the baseline flow exhibits periodic vortex shedding characteristic of a von Kármán vortex street, while the controlled conditions result in effective suppression of vortex shedding, leading to a streamlined wake and enhanced flow stability. In contrast, the controlled flow for $ Ar $ = 0.5 and $ Ar $ = 0.25 exhibits less effective suppression of vortex shedding. The controlled flow shows decreasing elongation of the recirculation zone and the wake transitions to a more intense multi-stage vortex shedding regime, where periodic oscillations persist despite the presence of active flow control.

In figure~\ref{fig:figure10}, we present the PSD results for the elliptical cylinders with four different $Ar$.
For the elliptical cylinders with $ Ar = 1 $ and $ Ar = 0.75 $, a comparative analysis of the PSD of the drag coefficient between the controlled and baseline flows indicates that the DRL-based active flow control strategy effectively modifies the spectral characteristics of the drag coefficient, completely suppressing the periodic vortex shedding behavior.
However, for the elliptical cylinders with $ Ar $ = 0.5 and $ Ar $ = 0.25, the PSD of the drag coefficient shows multiple peaks, reflecting the intricate wake dynamics caused by the extreme slenderness of the geometry. 
Although the control strategy reduces the PSD energy across the spectrum, it fails to fully suppress the primary peaks, leaving noticeable residual energy.

\begin{table*}[htb!]
\centering
\caption{Performance metrics of DRL-based flow control across various $Ar$.}
\label{tab:table1}
\begin{tabularx}{0.97\textwidth}{>{\centering\arraybackslash}p{1.2cm}
                                    >{\centering\arraybackslash}p{1.2cm}
                                    >{\centering\arraybackslash}p{1.3cm}
                                    >{\centering\arraybackslash}p{1.3cm}
                                    >{\centering\arraybackslash}p{3.4cm}
                                    >{\centering\arraybackslash}p{2cm}
                                    >{\centering\arraybackslash}p{2cm}}
\toprule
\toprule
$Ar$ & $Re$ & $C_\text{D, CFD}$ & $C_\text{D, Mean}$ & Drag reduction ratio (\%) & $C_\text{L, Mean}$ & $Q_\text{i,Mean}$\\ 
\hline
1.0  & 100 & 1.494 & 1.294 & 13.41 & 0.041 & -0.042 \\
0.75 & 100 & 1.696 & 1.356 & 20.08 & 0.007  & -0.007 \\
0.50  & 100 & 2.047 & 1.497 & 26.88 & -0.005 & 0.006 \\
0.25 & 100 & 2.593 & 1.806 & 30.37 & 0.107 & -0.164 \\
\bottomrule
\bottomrule
\end{tabularx}
\end{table*}

Table~\ref{tab:table1} summarizes the performance of the DRL-based AFC strategy for elliptical cylinders, demonstrating a significant reduction in the drag coefficient across all aspect ratios compared to the baseline flow. The effectiveness of the control strategy becomes more pronounced as the aspect ratio decreases.  
For the elliptical cylinder with $ Ar = 1 $, the DRL-based control achieves a 13.41\% reduction in drag. As the aspect ratio decreases, the drag reduction improves, reaching 20.08\% for $ Ar = 0.75 $ and 26.88\% for $ Ar = 0.5 $. In the most slender configuration ($ Ar = 0.25 $), the control strategy yields the highest drag reduction of 30.37\%. These results highlight the increasing effectiveness of DRL-based flow control in mitigating drag for more slender elliptical cylinders.
On the other hand, the external energy required to maintain control performance is represented by the mean value of the control actions. Across all aspect ratios of the elliptical cylinders, the control force remains at a relatively low level, demonstrating the energy efficiency of the control strategy. The DRL-based control strategy achieves effective flow regulation with minimal energy expenditure, further highlighting its efficiency.

Overall, the DRL-based flow control strategy exhibits strong effectiveness for elliptical cylinders with higher aspect ratios, such as $ Ar = 1 $ and $ Ar = 0.75 $, where it achieves near-complete suppression of dominant frequency peaks and substantial stabilization of the wake. However, as the aspect ratio decreases, the wake dynamics become increasingly complex, presenting greater challenges for the control strategy and leading to reduced performance in more slender configurations, particularly for aspect ratios of 0.5 and 0.25. Despite achieving considerable drag reductions of 26.88\% and 30.37\% for aspect ratios of 0.5 and 0.25, respectively, the persistence of residual energy at lower frequencies in these slender geometries underscores the need for further refinements to enhance control efficacy under extreme aspect ratio conditions.

\rewrite{
In the study by Zhang et al. \cite{he2023policy}, a DRL-based AFC strategy using the PPO algorithm effectively suppressed flow separation in a 2D setting at $Re = 100$. The control policy delayed separation on both sides of the cylinder by minimizing low-energy fluid near the separation region, improving overall flow behavior. This performance is consistent with the results obtained in the present study for a cylinder with $Ar = 1$ at the same Reynolds number.
Zhang et al. \cite{he2023policy} further applied the 2D-trained policy to three-dimensional (3D) flows with various spanwise lengths. The policy retained strong control capability and achieved notable drag reduction across all cases, despite variations in spanwise effectiveness. These findings confirm the generalizability and robustness of DRL policies trained in simplified 2D settings when applied to more realistic and complex 3D flow scenarios.}

\subsection{Suppression of vortex shedding in extremely slender elliptical cylinders}\label{sec:sec2}

In the previous section, we investigated the flow control performance of elliptical cylinders with aspect ratios $ Ar = 1 $, $ 0.75 $, $ 0.5 $, and $ 0.25 $. The results indicated that as the aspect ratio decreases, achieving complete suppression of vortex shedding becomes increasingly challenging, particularly in cases involving multi-stage shedding transitions.  
Building on these insights, we further evaluate our model under more stringent conditions in this section by applying it to a highly slender elliptical cylinder with $ Ar = 0.1 $. Additionally, we explore various approaches to enhance the fundamental DRL framework for more robust control performance.

As outlined in section~\ref{Enhanced}, we explore three different types of state representations and evaluate their impact on control performance. We start by employing the one-dimensional state formulation used in the previous subsection and apply it to the elliptical cylinder at $ Ar = 0.1 $. This formulation, denoted as \textit{Case A}, takes the pressure field at the final timestep of each control period as the characteristic state input. Building upon \textit{Case A}, we introduce \textit{Case B}, where the state representation is extended from a one-dimensional to a two-dimensional formulation. Specifically, we incorporate temporal information to enrich the state space, as described in \ref{Enhanced}. 
Instead of relying solely on the pressure value at the final timestep, we sample the pressure field across multiple timesteps within each control period. Out of the 50 timesteps per control cycle, 25 are selected to construct a temporal sequence representation, preserving richer spatiotemporal dynamics.

Lastly, in \textit{Case C}, we enhance the state estimation by integrating estimated pressure field obtained from Kalman-filtering. As evidenced by control results in the previous subsection, the flow is highly transient during the control process. Kalman filtering helps to provide potentially more accurate estimates for sparse data with less noise. By combining raw observations with KF-predicted states, we aim to improve the quality of data from the limited number of probes to construct a more robust state representation, facilitating the agent’s ability to capture the evolving flow dynamics.

\begin{figure}
    \includegraphics[width=\textwidth]{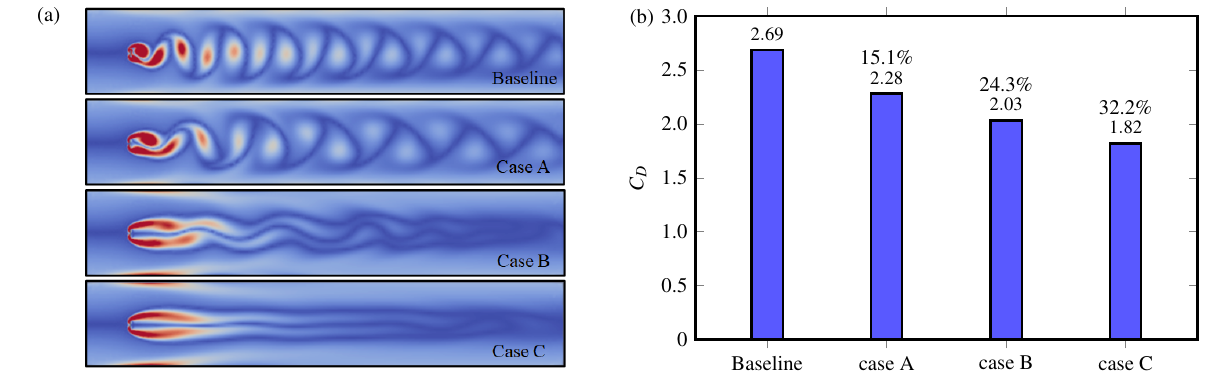}
    \caption{  
    Comparison of control performance for the elliptical cylinder with $ Ar = 0.1 $ under various cases. (a) Vorticity contours of the baseline and controlled flows; (b) Drag reduction rate under different configurations.
    }
\label{fig:figure11}
\end{figure}

The flow control performance results under the three cases are illustrated in figure~\ref{fig:figure11}.  
In \textit{Case A}, the control strategy mitigates vortex shedding in the wake of the elliptical cylinder to a certain extent, achieving a 15.1\% reduction in drag compared to the baseline flow. However, periodic vortex structures persist in the wake, indicating the limited effectiveness of this control scheme in fully suppressing vortex shedding. 
This is in agreement with the findings from previous results where geometries with lower $Ar$ are more challenging to control.
In contrast, with the help of temporal information, the control model in \textit{Case B} significantly enhances flow control performance by further suppressing vortex shedding and reducing the drag coefficient by 24.3\%. However, there is still observable oscillatory vortex shedding in the flow. We have attempted to further improve the model with extensive hyperparameter tuning and training, but failed to obtain a completely stable flow field.

The model in \textit{Case C} which integrates Kalman-filtered data achieves the optimal control effect, reducing the drag coefficient by 32.2\% while completely suppressing vortex shedding in the wake of the elliptical cylinder, demonstrating the most effective flow control capability. 
We conjecture that the predicted values from Kalman filter are of higher quality and complement the raw pressure field data, allowing the DRL agent to learn a more effective control strategy. Overall, we find that enriching the state space with additional data, such as temporal history and less noisy estimates could tremendously improve the performance of DRL agent, even achieving full vortex suppression for an extremely slender geometry at $Ar = 0.1$.

\begin{figure}
\includegraphics[width=0.97\textwidth]{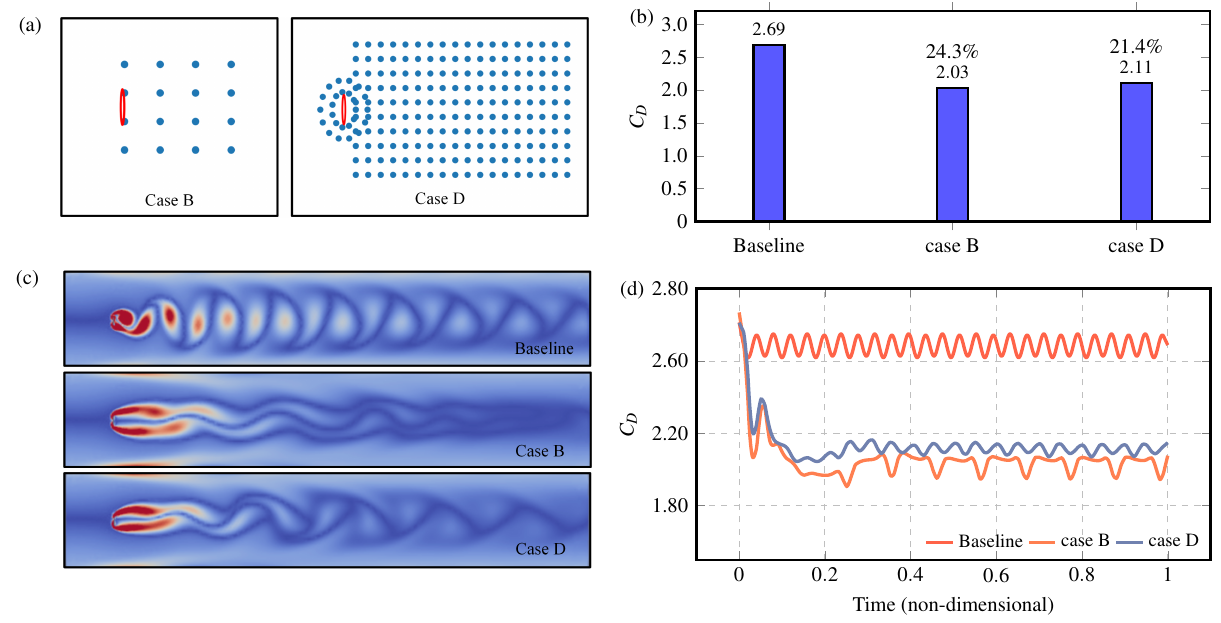}
\caption{(a) Two probe distributions; (b) Drag reduction performance; (c) Vortex suppression effect; (d) Drag coefficient of the elliptical cylinder.}  
\label{fig:figure12}
\end{figure}

In addition to enriching the state space with temporal information, we would like to investigate whether modifications to the spatial probe distribution can enhance control performance as well. To this end, we design \textit{Case D} by densifying the probe distribution while maintaining temporal resolution, similar to the state representation used in \textit{Case B}. 
Additional probes are strategically placed around the elliptical cylinder and within the wake region, enabling a more comprehensive extraction of flow field information.  
Figure~\ref{fig:figure12} presents the probe distributions for both cases along with their respective control performances.  
Although the drag coefficient of the elliptical cylinder is reduced by 21.4\% in \textit{Case D}, the suppression of vortex shedding is less effective compared to that in \textit{Case B}. 
Surprisingly, we find that densifying the probe distribution actually weakens the control effectiveness. 
We note that optimization of model hyper-parameters failed to yield any substantial improvement in model performance. 
This observation suggests that the probe configuration in \textit{Case B} is sufficiently effective, and additional probes for flow field extraction does not necessarily yield further performance improvements due to over-fitting.

Moreover, the controlled flow in both \textit{Case B} and \textit{Case D} still exhibits significant fluctuations in the drag coefficient. The wake of the controlled flow presents a complex vortex shedding pattern and pronounced wake instabilities. These observations highlight the considerable challenge of achieving complete vortex shedding suppression in highly slender body flow scenarios.

\begin{figure}[hbt!]
    \centering
    \includegraphics[width=\textwidth]{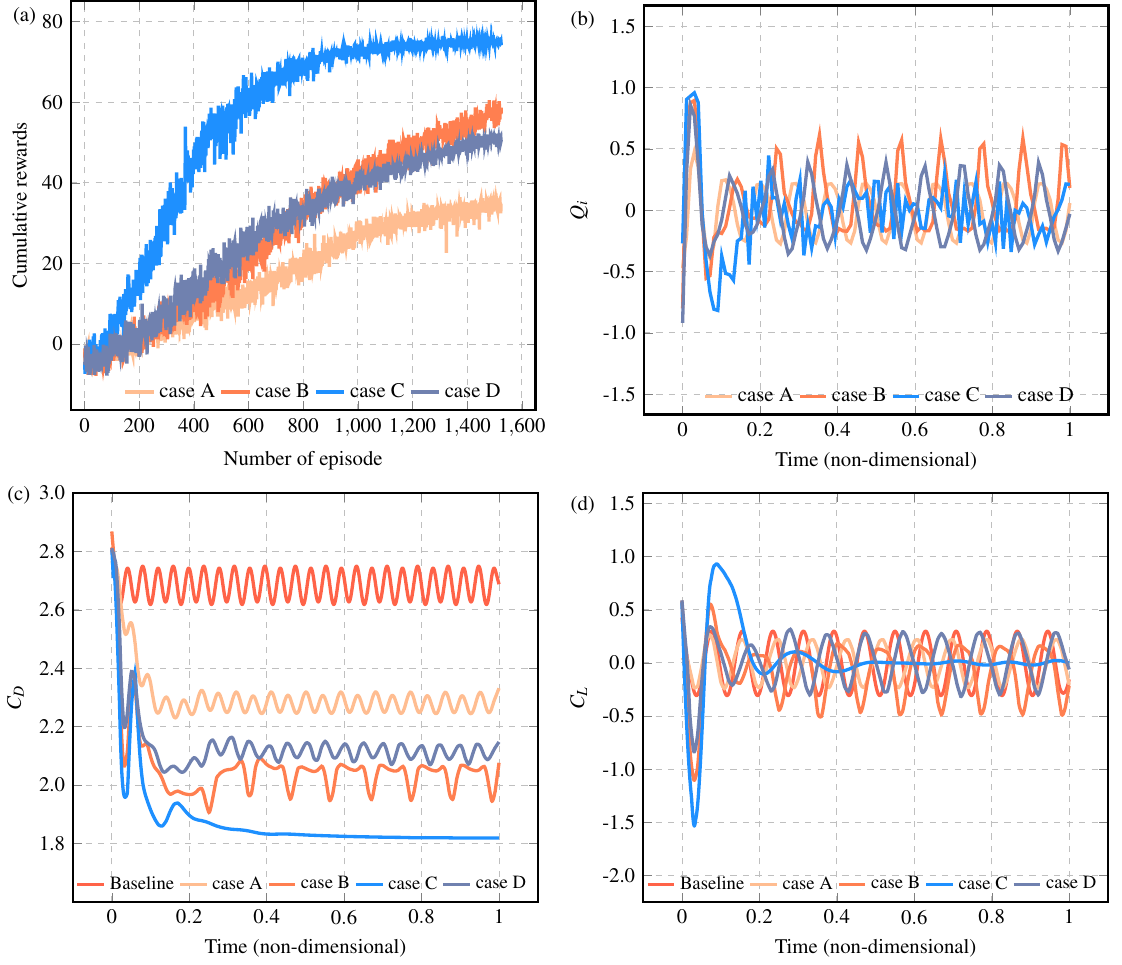}
    \caption{  
    Comparison of control performance for the elliptical cylinder with $ Ar = 0.1 $ under various cases. (a) Reward function; (b) the mass flow rate of the synthetic jet; (c) Lift coefficient ;(d) drag coefficient.  
    }
\label{fig:figure13}
\end{figure}

\begin{figure}[hbt!]
\centering
    \includegraphics[width=\textwidth]{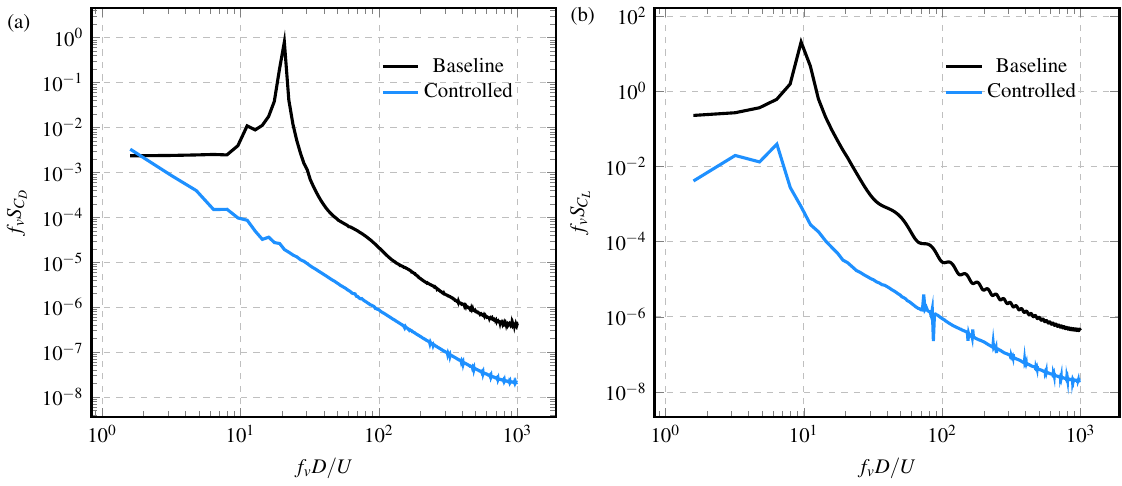}
    \caption{Comparison of temporal and spectral characteristics of $C_D$ and $C_L$ for baseline and controlled flows under case C. (a) power spectral densities of $C_D$; (b) power spectral densities of $C_L$.}
\label{fig:figure14}
\end{figure}

Figure~\ref{fig:figure13} compares the training history and control performance of DRL agent under the four different cases for $Ar = 0.1$, illustrating the evolution of the reward function, synthetic jet mass flow rate, and aerodynamic force coefficients during the DRL training process.  
Among the four cases, \textit{Case C} demonstrates the most effective control strategy. It achieves the highest cumulative reward and the fastest convergence rate, indicating superior learning efficiency and control precision.  
\textit{Case C} nearly eliminates the periodic oscillations of the lift coefficient, stabilizing it close to zero. This represents a significant improvement compared to the other cases, where fluctuations persist. Moreover, while the synthetic jet mass flow rate is initially high at the onset of control, \textit{Case C} exhibits a gradual and sustained decline, reflecting an adaptive and energy-efficient control mechanism.  
These results collectively highlight the effectiveness and robustness of the control strategy in \textit{Case C}.

To further substantiate our findings, we plot the power spectral densities for the baseline flow and controlled flow in \textit{Case C} in figure~\ref{fig:figure14}. 
Prominent peaks at dominant frequencies are evident in the baseline flow. These peaks are effectively suppressed in the controlled flow, leading to substantial redistribution of spectral energy and a broadband reduction in energy content. 
The findings indicate that integrating Kalman-filtered state information significantly enhances control performance.
The DRL agent which previously struggled with low $Ar$ values could ultimately learn an optimal control strategy for the extremely elliptical cylinder, with the help of an appropriately enriched state space. The optimized control strategy not only achieves complete suppression of vortex shedding but also minimizes drag while eliminating periodic fluctuations in the drag coefficient, ultimately leading to a fully stabilized wake. 

\begin{figure}
\centering
\includegraphics[width=0.75\textwidth]{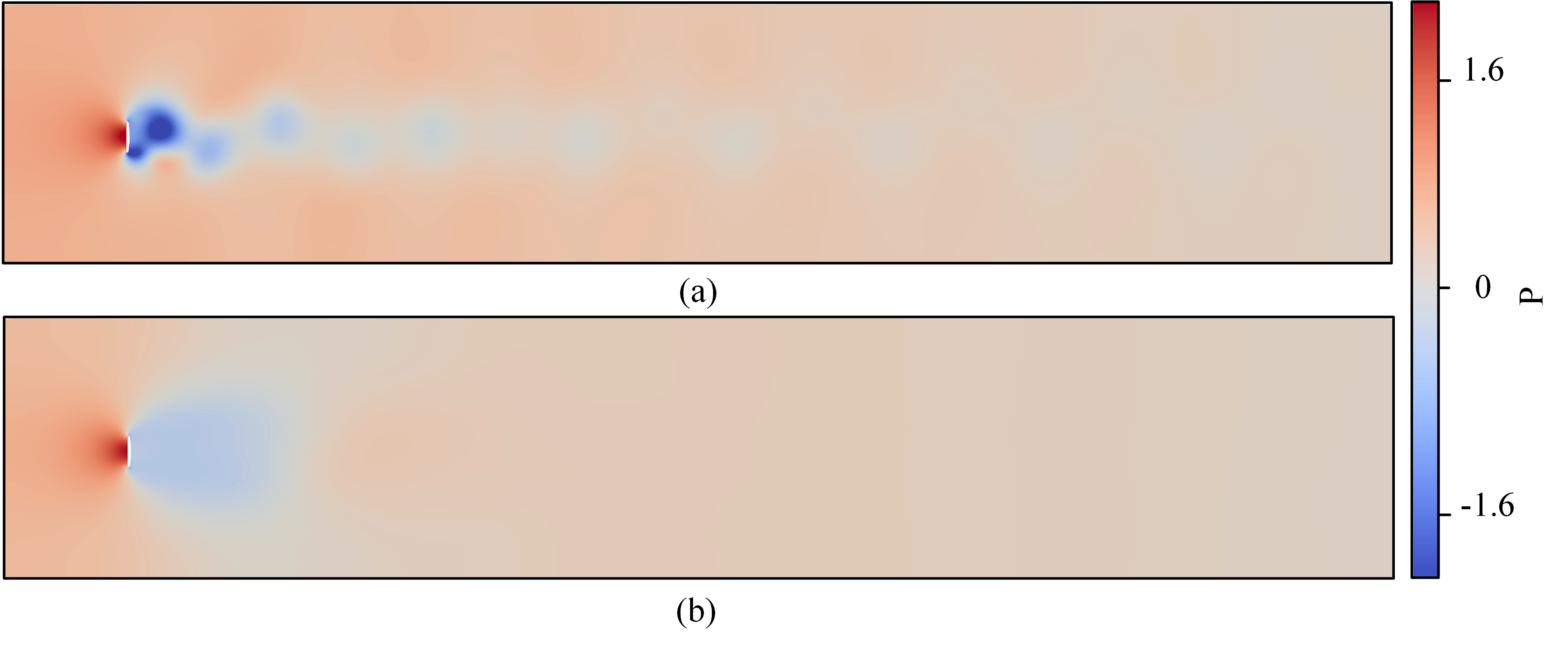}
\caption{Flow structures around the elliptical cylinder. Pressure fields: (a) baseline flow; (b) controlled flow under case C.}
\label{fig:figure15}
\end{figure}

Similarly, the pressure field around the elliptical cylinder shown in figure \ref{fig:figure15} also illustrates the effectiveness of the control strategy.
The pressure field under baseline conditions reveals alternating high- and low-pressure regions downstream, resulting in unsteady aerodynamic forces. 
The control strategy minimizes these pressure variations, leading to a more uniform pressure distribution and a reduction in unsteady forces acting on the cylinder. 
These findings highlight the adaptability of DRL-based AFC strategy. By enhancing the DRL framework with enriched states and optimal probe location, the DRL agent is capable of achieving the desirable objective of vortex suppression for a highly slender cylinder with $Ar = 0.1$.

\begin{figure}
    \centering
    \includegraphics[width=0.75\textwidth]{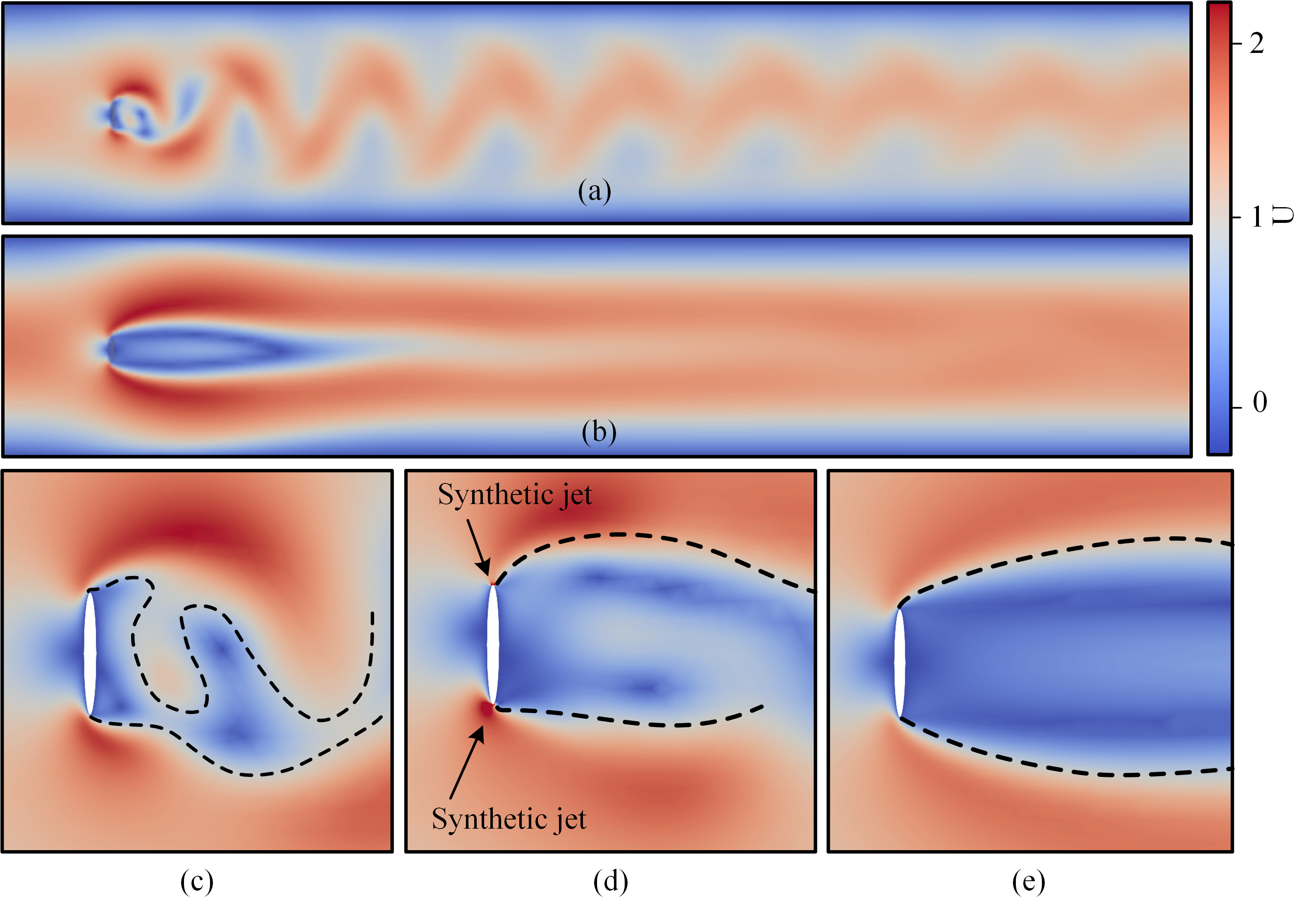}
    \caption{(a) represents the baseline flow for $ Ar = 0.1 $, (b) depicts the controlled flow under case C, (c) illustrates the flow structure around the elliptical cylinder under baseline conditions, (d) provides a detailed view of the flow around the elliptical cylinder during the control process, and (e) presents the detailed flow structure around the elliptical cylinder after the control has been applied.}
\label{fig:figure16}
\end{figure}

To provide a more visual illustration of how the DRL-based control strategy suppresses vortex shedding at $ Ar = 0.1 $, figure~\ref{fig:figure16} presents both the global flow field and the detailed flow structures surrounding the elliptical cylinder.
In the baseline flow depicted in figure \ref{fig:figure16} (a), alternating vortex shedding manifests a complex multi-stage structure characterized by prominent recirculation zones and significant fluctuations downstream of the elliptical cylinder.
In contrast, figure \ref{fig:figure16} (b) presents the controlled flow, where periodic vortex shedding is completely suppressed. This suppression results in a streamlined wake with an expanded recirculation region , which signifies a transition to a more stable flow regime.
Figure \ref{fig:figure16} (c)-(e) further illustrates close-up views near the highly slender cylinder. Before any flow control, the baseline flow is characterized by distinct separation regions along both sides of the cylinder, contributing to wake instability. 

During the initial stages of control, as shown in figure \ref{fig:figure16} (d), a gradual stabilization of the recirculation region is observed. This is evidenced by a notable reduction in the intensity of near-wake vortex structures, indicating the onset of effective flow control.
By the end of the control process, illustrated in figure \ref{fig:figure16} (e), the flow achieves full stabilization. In this final state, the wake transitions to a smooth, elongated structure devoid of coherent vortex shedding. This transformation underscores the efficacy of the control strategy in managing complex wake dynamics and attaining optimal flow control objectives.
In summary, the progressive stabilization observed throughout these phases highlights the control strategy's capability to transform unstable flow conditions into a well-ordered state, significantly enhancing the aerodynamic performance around the elliptical cylinder.

\section{CONCLUSIONS}\label{sec:Conclusions} 

We propose a deep reinforcement learning-based framework for active flow control, leveraging a PPO agent to dynamically regulate the mass flow rates of two synthetic jets on the surface of elliptical cylinders, with varying $Ar$. 
This approach effectively suppresses vortex shedding, enhances wake stability, and achieves substantial drag reduction. Notably, it maintains robust performance even for highly slender elliptical cylinders, where flow control becomes increasingly challenging.  
The primary contributions of this study are summarized as follows:  

\begin{enumerate}
\item \textbf{Flow control effectiveness for elliptical cylinders with $Ar=(1.0, 0.75, 0.5, 0.25)$:}  
The flow control strategy trained using deep reinforcement learning demonstrates varying degrees of effectiveness across elliptical cylinders with different aspect ratios. For elliptical cylinders with higher aspect ratios ($Ar = 1$ and $Ar = 0.75$), the strategy effectively mitigates vortex shedding, achieving near-complete suppression of dominant frequency peaks and resulting in drag reductions of 13.41\% and 20.08\%, respectively. However, as the aspect ratio decreases, the performance of the flow control strategy gradually diminishes where the drag reduction ratio increases but the controlled flow becomes increasingly oscillatory and unstable. For an elliptical cylinder with $Ar = 0.5$, the strategy achieves only partial suppression of vortex shedding, with a drag reduction of 26.88\%. In the case of $Ar = 0.25$, the strategy achieves the highest drag reduction of 30.37\%, but it struggles to fully suppress dominant frequency peaks. This highlights the challenges associated with implementing flow control in bluff-body flows around highly slender elliptical geometries, where wake dynamics become increasingly complex with decreasing values of $Ar$.

\item \textbf{Suppression of vortex shedding for an extremely slender cylinder at $Ar = 0.1$:}  
For a highly slender elliptical cylinder ($Ar = 0.1$), the effectiveness of the basic DRL-based flow control framework is significantly limited. To address this, we enhanced the control framework by enriching the state space with temporal information and incorporating Kalman filtering to provide accurate estimates with reduced noise. This improved strategy successfully achieved complete suppression of vortex shedding, with a maximum drag reduction rate of 32.2\%, and fully stabilized the wake flow around the elliptical cylinder at $Ar = 0.1$. Furthermore, we investigated the adverse impact of employing an excessive number of probes, which leads to over-fitting. 
These results highlight the critical role of optimizing probe distribution and appropriately enhancing state representations in effectively managing wake dynamics for slender geometries. 
With a balanced state space for the DRL agent, the optimized control strategy simultaneously achieves multiple control objectives for the extremely slender cylinder at $Ar = 0.1$, including significant reductions in drag and lift coefficients, elimination of periodic oscillations in drag and lift, and stabilization of the flow field.

\end{enumerate}

In conclusion, this study addresses critical challenges associated with the placement of observation sensors and the extraction of state information, thereby significantly enhancing the control performance of DRL-based AFC. The proposed framework demonstrates improved robustness and efficiency in suppressing vortex shedding over a range of elliptical geometries. However, the current investigation is limited to 2D flows at relatively low Reynolds numbers, which may not fully capture the complexities of real-world 3D turbulent wakes. Future research will aim to extend this approach to higher Reynolds number regimes and fully 3D configurations, incorporating more realistic flow features and control constraints to further bridge the gap between numerical studies and practical applications.

\section{APPENDIX I}\label{app1}

\rewrite{The numerical simulations are performed using a transient incompressible flow solver based on the \texttt{OpenFOAM} platform. The governing equations are discretized using the finite volume method, and the velocity–pressure coupling is handled via the PIMPLE algorithm, which combines features of both the PISO and SIMPLE methods for improved stability in unsteady flow computations.
Table~\ref{tab:table2} summarizes the spatial and temporal discretization schemes adopted in this study.}

\begin{table}[htbp]
\centering
\caption{Spatial and temporal discretization schemes used in numerical simulation.}
\label{tab:table2}
\begin{tabular}{p{0.35\textwidth}p{0.55\textwidth}}
\toprule
\toprule
timeSchemes       & Backward \\
\hline
GradSchemes       & Gauss linear \\
\multirow{3}{=}{divSchemes} 
                  & div(phi, U) \hfill Gauss linear \\
                  &             \hfill Upwind grad(U) \\
                  & div((nuEff* dev2(T(grad(U))))) \hfill Gauss linear \\
snGradSchemes     & Corrected \\
laplacianSchemes  & Gauss linear corrected \\
\bottomrule
\bottomrule
\end{tabular}
\end{table}

\begin{figure*}[hbt!]
    \centering
    \includegraphics[width=0.8\textwidth]{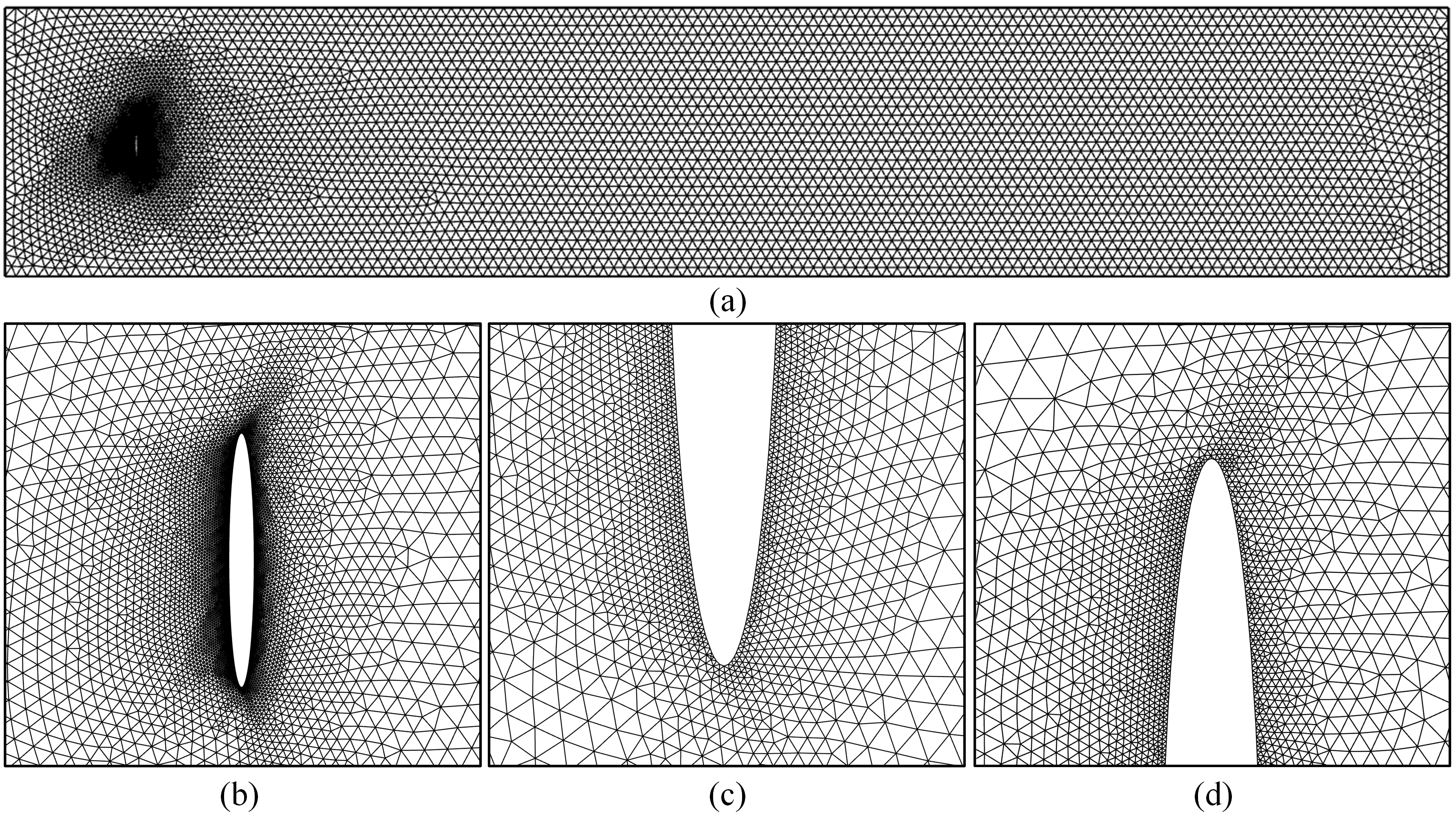}
    \caption{Discretization of the computational domain for the cylinder with $Ar$ = 0.1. (a) Global mesh of the computational domain. (b) Mesh around the cylinder. (c) and (d) Part of the mesh on the cylinder.}
    \label{fig:figure17}
\end{figure*}

\rewrite{To ensure the accuracy and reliability of the numerical simulations, a grid independence study is performed for an elliptical cylinder with an aspect ratio of $Ar = 0.1$ at a Reynolds number of $Re = 100$. The computational domain is discretized using unstructured mesh elements, as illustrated in figure~\ref{fig:figure17}. To assess mesh sensitivity, three grid resolutions—coarse, medium, and fine—are examined, with their corresponding mesh parameters and key flow characteristics summarized in table~\ref{tab:table3}.
The simulation results reveal that the medium and fine meshes produce nearly identical force coefficients and vortex shedding frequencies, indicating convergence of the solution. In contrast, the coarse mesh exhibits significant deviations, particularly in the predicted lift and drag forces. Based on this comparison, the medium mesh is selected for all subsequent simulations, as it provides a satisfactory balance between numerical accuracy and computational efficiency.
Moreover, similar grid independence tests are conducted for other aspect ratios considered in this study to ensure the consistency and robustness of the numerical results across different geometric configurations.}

\begin{table*}[hbt!]
\centering
\caption{Grid independence test for a cylinder with $Ar$ = 0.1 at $Re$ = 100.}
\begin{tabular}{
  >{\centering\arraybackslash}p{0.12\textwidth}
  >{\centering\arraybackslash}p{0.15\textwidth}
  >{\centering\arraybackslash}p{0.1\textwidth}
  >{\centering\arraybackslash}p{0.1\textwidth}
  >{\centering\arraybackslash}p{0.1\textwidth}
  >{\centering\arraybackslash}p{0.1\textwidth}
  >{\centering\arraybackslash}p{0.1\textwidth} 
}
\toprule
\toprule
Configuration & Resolution & Mesh  & $C_\text{D, Max}$ & $C_\text{D, Mean}$ & $C_\text{L, Max}$ & $St$ \\
\hline
        & {Fine}     & 28,142 & 2.672 & 2.657 & 0.301  & 0.497 \\
$Ar = 0.1$  & {Medium}   & 22,142 & 2.784 & 2.683 & 0.312  & 0.502 \\
        & {Coarse}   & 14,424 & 2.842 & 2.722 & 0.427  & 0.542 \\
\bottomrule
\bottomrule
\end{tabular}
\label{tab:table3}
\end{table*}

\section{APPENDIX II}\label{app2}

In the problem of state estimation for dynamic systems, the Kalman Filter serves as an optimal recursive estimation algorithm, effectively extracting the true state of the system in high-noise environments. This study employs a Real-Time Kalman Filter to process time-series data and further constructs an \textit{Enhanced State} to improve the representation capability of system state information.
The Kalman Filter estimate system states by dynamically adjusting the estimated values based on current observations, minimizing uncertainty \cite{6400245}. The core process of the Kalman Filter can be described by the following equation:
\begin{equation}
    \hat{\mathbf{x}}_t = \mathbf{A} \hat{\mathbf{x}}_{t-1} + \mathbf{K}_t (\mathbf{z}_t - \mathbf{H} \mathbf{A} \hat{\mathbf{x}}_{t-1})
\end{equation}
where $ \hat{\mathbf{x}}_t $ is the estimated state at time $ t $, evolving through the state transition matrix $ \mathbf{A} $ from $ \hat{\mathbf{x}}_{t-1} $. The Kalman gain $ \mathbf{K}_t $ adjusts the predicted state using the observation $ \mathbf{z}_t $, which is mapped from the state space via the observation matrix $ \mathbf{H} $.
The Kalman gain $\mathbf{K}_t$ is given by:
\begin{equation}
    \mathbf{K}_t = \mathbf{P}_t^{-} \mathbf{H}^T (\mathbf{H} \mathbf{P}_t^{-} \mathbf{H}^T + \mathbf{R})^{-1}
\end{equation}
where the predicted uncertainty $ \mathbf{P}_{t}^{-} $ is large and the observation noise $ \mathbf{R} $ is small, the Kalman gain $ \mathbf{K}_t $ increases, making the update rely more on observations. Conversely, if $ \mathbf{R} $ is large, $ \mathbf{K}_t $ decreases, and the update depends more on the prediction.
This expression incorporates both the prediction step $\mathbf{A} \hat{\mathbf{x}}_{t-1}$ and the update step (correction based on the observation $\mathbf{z}_t$), forming the fundamental computational procedure of the Kalman Filter.

\section*{Conflict of Interest}

The authors report no conflict of interest.

\bibliographystyle{elsarticle-num-names} 
\bibliography{elsarticle}

\end{document}